\documentclass[letterpaper,11pt]{rsauthor}
\jname{Proc. R. Soc. A}
\begin{document}

\title{AARTFAAC: Towards a 24x7, All-sky Monitor for LOFAR}
\author{Peeyush Prasad$^1$ and Stefan J. Wijnholds$^2$}
\address{$^1$Anton Pannekoek Institute, Universiteit Van Amsterdam, The Netherlands\\ $^2$Netherlands Institute for Radio Astronomy (ASTRON), Oude
  Hoogeveensedijk 4, 7991 PD Dwingeloo, The Netherlands}
\date{\today}

\abstract{
  The AARTFAAC project aims to implement an All-Sky Monitor (ASM),
  using the Low Frequency Array (LOFAR) telescope. It will enable
  real-time, 24x7 monitoring for low frequency radio transients over
  most of the sky locally visible to the LOFAR at timescales ranging
  from milliseconds to several days, and rapid triggering of follow-up
  observations with the full LOFAR on detection of potential transient
  candidates. These requirements pose several implementation
  challenges: imaging of an all-sky field of view, low latencies of
  processing, continuous availability and autonomous operation of the
  ASM. The first of these has already resulted in the correlator for
  the ASM being the largest in the world in terms of its number of
  input channels. It will generate $\sim 1.5 \cdot 10^5$ correlations
  per second per spectral channel when built. Test observations using
  existing LOFAR infrastructure were carried out to quantify and
  constrain crucial instrumental design criteria for the ASM. In this
  paper, we present an overview of the AARTFAAC data processing
  pipeline and illustrate some of the aforementioned challenges by
  showing all-sky images obtained from one of the test
  observations. These results provide quantitative estimates of the
  capabilities of the instrument.
}


\maketitle

\section{Introduction}

The recent serendipitous discoveries of several astrophysical radio
transients at a variety of flux and time scales (see
e.g. \cite{Cordes2007SKAmem}) has opened up a new window in the search
for exotic objects of both known and unknown type. It is felt that
the transient or dynamic radio sky, specially at low frequencies,
should be rich enough to benefit from blind surveys along the lines of
wide field instruments at higher energies (X and $\gamma$ rays), which have
been very successful at detecting transient sources. Further, it is
argued that for the problem of transient detection, one is better off
having a much larger field of view, while trading off sensitivity
\cite{Cordes2007SKAmem}.
 
It is in this context that the Amsterdam-ASTRON Radio Transient
Facility And Analysis Center (AARTFAAC), a collaboration between
ASTRON and the University of Amsterdam, aims to implement a near
real-time, 24x7 All-Sky Monitor (ASM) for the LOFAR. 
Such an instrument will enable monitoring for
low frequency radio transients over most of the sky locally visible to
the LOFAR at timescales ranging from seconds to several days. In the
following sections, we introduce the ASM in more detail and present
some initial results.

\section{The AARTFAAC ASM}

\begin{figure}
\centering
\includegraphics[width=\textwidth]{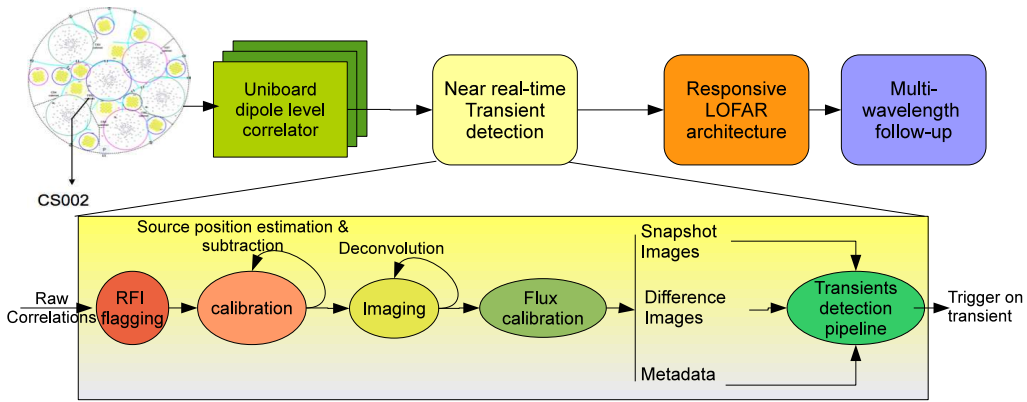}
\caption{The main components of the AARTFAAC ASM. \label{fig:blkdiag}}
\end{figure}

LOFAR \cite{Vos2009ProcIEEE} is an array of 'stations' spread over
hundreds of kilometers, while each station is itself an array composed
of two kinds of receiving elements: dipoles or Low Band Antennas (LBA)
operating between 10 -- 80 MHz, and tiles or High Band Antennas (HBA),
operating between 110 -- 240 MHz. The ASM will use six stations at the
heart of LOFAR and will be a zenith-pointing, transit mode
instrument. This configuration makes the ASM a 288-element array,
spread over $\sim$350 m providing almost full UV coverage and a well
defined PSF which does not change with time.  Usually, the digitized
outputs of every element in a station are coherently summed to form a
beam in a given direction before being available for
correlation. This, however, restricts the field of view of a station
to a few degrees. The ASM needs to correlate the signals from all
signal paths to image the full field of view ($2\pi$ sr (all-sky) for
LBA and $\sim$1.5 -- 10\% of the sky for the HBA, depending on the
observing frequency). For continuous monitoring, the ASM operates in a
piggyback fashion simultaneously with regularly scheduled LOFAR
observations, thus sharing their observational parameters. The overall
control flow and main components of the ASM are depicted in Fig.\
\ref{fig:blkdiag}, and are further described below.

The AARTFAAC correlator will have 576 inputs (dual polarization from
288 elements) leading to $\sim 1.5 \cdot 10^5$ correlations for each
spectral channel. Its implementation scheme is described in
\cite{Gunst2011ASTRON}. The current hardware specifications assume 24
kHz spectral and 1 second temporal resolution (to prevent time and
bandwidth smearing). This results in a sensitivity of $\sim$4 Jy at 60
MHz. The resolution of 0.8 square degrees leads to a confusion noise
of $\sim$8 Jy at 60 MHz, leading to the expectation of the ASM being
confusion noise, rather than thermal noise limited. The total
available bandwidth is $\sim$13 MHz, which can be arbitrarily
distributed over the 100 MHz total available digitized band. Thus, the
ASM has a very versatile instantaneous spectral coverage over two
octaves in the LBA, and one octave in HBA.

The correlator outputs are first passed through an RFI excision stage,
the challenge being to generate appropriate RFI masking with the
limited temporal and spectral baseline available due to the near
real-time nature of the system. The correlations are then calibrated
and imaged with low latency. This stage may require multiple
iterations and dominates the computing. The output images will then be
flux calibrated, forming the input to the Transients Pipeline \cite{Swinbank2007PoS}.
This carries out source extraction, source association, and the
generation of light curves from existing observations for transient
detection. It will also generate low-latency triggers to a LOFAR
architectural module termed the 'Responsive LOFAR module', which can
trigger multi-wavelength follow-up observations with a variety of
instruments.

\section{ASM Calibration Challenges}

ASM calibration refers to the estimation of a complex direction
independent gain per antenna and direction dependent parameters per
calibration source, characterizing the ASM and propagation through the
ionosphere. Traditionally, calibration is carried out periodically,
with the expectation that instrumental and observational parameters
remain stable in the interim. However, the ionosphere can have a
significant effect on the propagation of low frequency radio waves as
observed by the ASM. Thus, the ASM requires continuous, direction
dependent calibration, i.e., calibration of each timeslice. We use a
Weighted Alternating Least Squares algorithm for multi-source
self-calibration, as described in \cite{Wijnholds2009TSP}. The
algorithm solves for the best fitting calibration solutions in a
Least-Squares sense and requires an initial sky model.

Transient detection is proposed via either analysis of each source's
light curve, generated by source extraction on every image timeslice,
or via image level differencing. Both techniques require the
minimizing of calibration errors to reduce the false detection rate of
transients. One of the contributors to calibration errors are shifts
in the observed positions of the sources in the sky-model due to
ionospheric refraction, leading to model visibilities differing from
observed visibilities, ultimately resulting in the non-convergence of
the calibration process. We address this by estimating source
positions from data using Weighted Subspace Fitting (WSF)
\cite{Viberg1991TSP}, which is incorporated into every calibration
cycle. The dynamic range of calibrated images is then improved by the
subtraction of the visibility contribution of the brightest sources in
the sky, along with their sidelobes.

\begin{figure}
\centering
\includegraphics[width=0.48\textwidth]{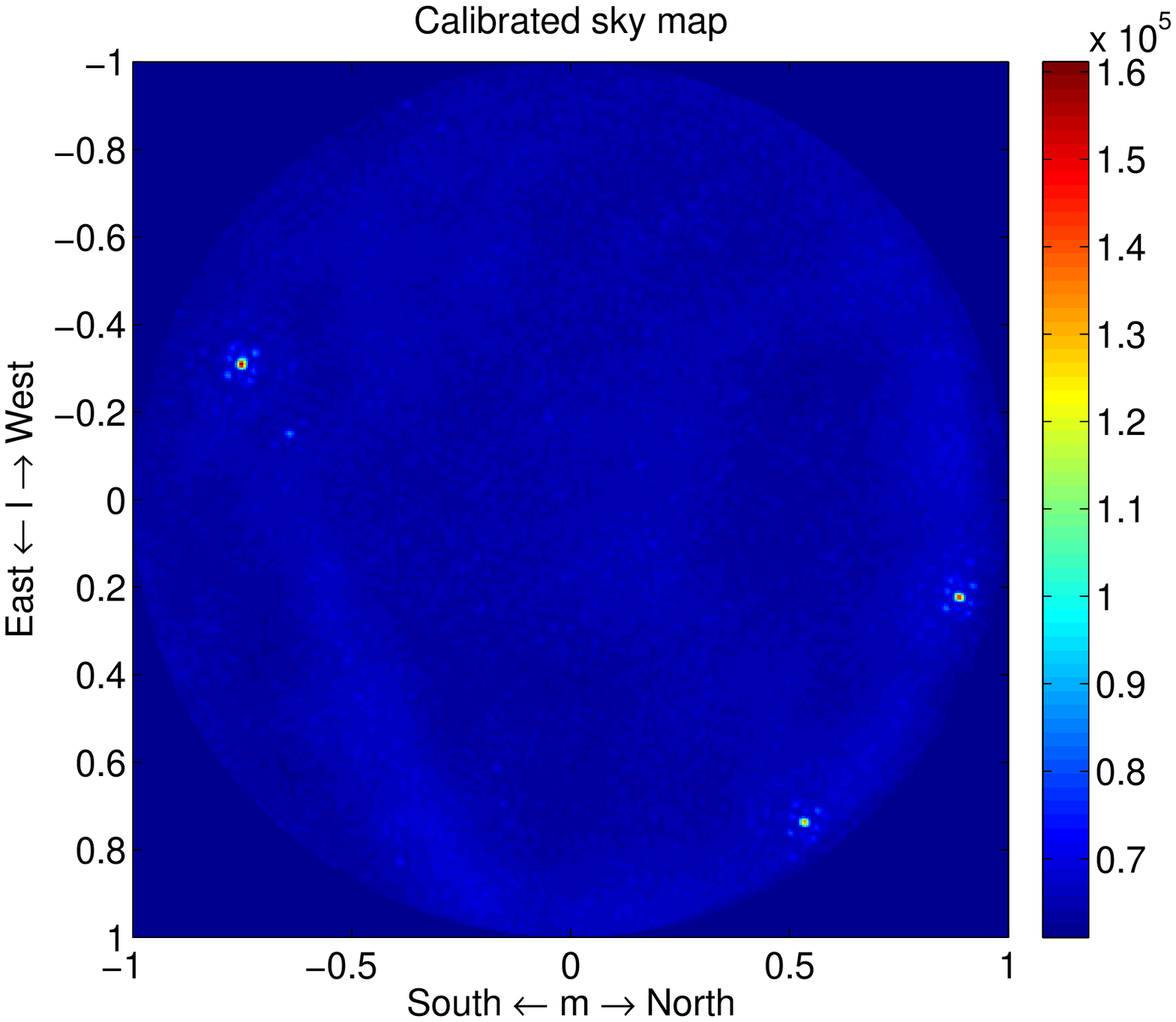}
\includegraphics[width=0.48\textwidth]{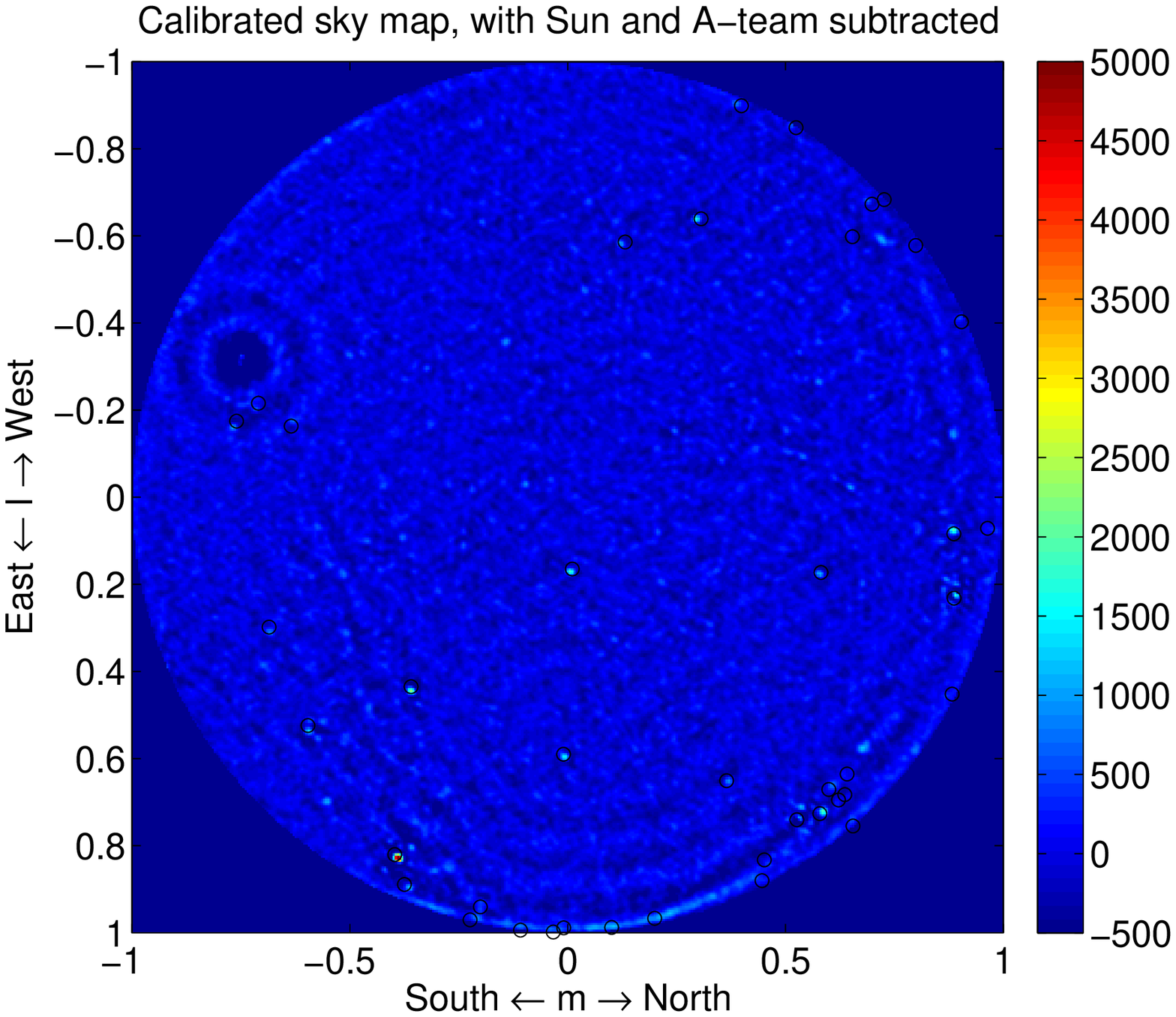}
\caption{: (Left) Calibrated skymap with flaring Sun, CasA, CygA and
  Galactic plane visible. (Right) Map with the Sun and bright sources
  subtracted and the Galactic plane filtered out, making weaker
  sources visibile. Over-plotted circles indicate the position of
  several 3C sources. \label{fig:allsky_image}}
\end{figure}

The ASM just resolves the Sun. This precludes its modeling as a point
source, while also preventing the suppression of the Solar flux by
eliminating short baselines. During solar flares the Sun can be the
dominant contributor of flux to visibilities, while morphing into a
source containing multiple complicated components which change over
the duration of the flare. This was discovered in test observations,
in which calibration sometimes failed during significant solar
activity. We have applied a sparse-reconstruction based algorithm to
estimate a reasonable model of the flaring Sun in an automated
manner. This was found to be effective, although being compute
intensive. The approaches described above allowed for effective removal of the
Sun and the bright radio sources dominating the observed
visibilities. For algorithm validation while hardware is being developed, test data from all dipoles of six stations was acquired using existing LOFAR hardware and software. Fig.\ \ref{fig:allsky_image} shows images
from data acquired on September 21, 2011, 12:39 hours UTC
before and after the removal of the Sun and bright sources. The noise
in this image, generated using an time integration of 10 s and
$\sim$90 kHz bandwidth, indicates a dynamic range of $\sim$2200:1.

\section{Conclusion}

The AARTFAAC ASM will be one of the first all-sky monitors at radio
wavelengths. Its calibration is challenging because of the dynamic
nature of the low frequency observations due to factors like an
active ionosphere over an extremely large field of view, or 
Solar activity. We have shown that advanced algorithms can effectively
address some factors at the cost of increased computing. The
ASM's implementation is ongoing, with appropriate hardware 
procured and firmware in development. Appropriate calibration
approaches are being developed based on experience gained via test
observations.


\begin{thebibliography}{1}

\bibitem{Cordes2007SKAmem}{[1]}
Jim Cordes. {The SKA as a Radio Synoptic Survey Telescope: Widefield Surveys for
  Transients, Pulsars and ETI}. {\em {SKA memo series}}, 97, September 2007.

\bibitem{Vos2009ProcIEEE}{[2]}
Marco de~Vos, Andre~W. Gunst, and Ronald Nijboer. {The LOFAR
  Telescope: System Architecture and Signal Processing}. {\em
  {Proceedings of the IEEE}}, 97(8):1431--1437, August 2009.

\bibitem{Gunst2011ASTRON}{[3]}
Andre~W. {Digital Correlator System for AARTFAAC}. Technical Report
{ASTRON-RP-331}, {ASTRON}, Dwingeloo, The Netherlands, 12 July 2011.

\bibitem{Swinbank2007PoS}{[4]}
John Swinbank et~al.  {A transient detection and monitoring pipeline
  for LOFAR}.  In {\em {Proceedings of Science}}, 12- 15June 2007.

\bibitem{Wijnholds2009TSP}{[5]}
Stefan~J. Wijnholds and Alle-Jan van~der Veen. {Multisource
  Self-calibrationfor Sensor Arrays}. {\em {IEEE Transactions on Signal Processing}}, 57(9):3512--3522, September 2009.

\bibitem{Viberg1991TSP}{[6]}
Mats Viberg, {Bj\"orn} Ottersten, and Thomas Kailath. {Detection and
  Estimation in Sensor Arrays Using Weighted Subspace Fitting}.  {\em
  {IEEE Transactions on Signal Processing}}, 39(11):2436--2449,
November 1991.

\end{thebibliography}
\end{document}